\documentstyle[12pt]{article}
\begin{document}
\def\thesection{\Roman{section}}
\thispagestyle{empty}
\begin{center}
\LARGE \tt \bf{Geodesics around gravitational dislocations}
\end{center}
\vspace{1cm}
\begin{center} {\large L.C. Garcia de Andrade\footnote{Departamento de
F\'{\i}sica Te\'{o}rica - Instituto de F\'{\i}sica - UERJ
Rua S\~{a}o Fco. Xavier 524, Rio de Janeiro, RJ
Maracan\~{a}, CEP:20550-003 , Brasil.
E-Mail.: GARCIA@SYMBCOMP.UERJ.BR}}
\end{center}
\vspace{1.0cm}
\begin{abstract}
A technique for generating spherically symmetric dislocation solutions of a direct Poincar\'{e} 
gauge theory of gravity based on homogeneous functions which makes Cartan torsion to vanish is
presented.Static space supported dislocation and time dependent solutions are supplied.Photons 
move along geodesics in analogy to geodesics described by electrons around dislocations in 
solid state physics.Tachyonic sectors are also found.
\end{abstract}      
\vspace{1.0cm}       
\begin{center}
\Large{PACS numbers : 0420,0450.}
\end{center}
\newpage
\pagestyle{myheadings}
\section{Introduction}
\paragraph*{}
In a series of papers D.G.B.Edelen \cite{1,2,3,4,5} produced a technique for generating defect solutions (dislocations and disclinations) of Einstein field equations of a direct
Poincar\'{e} gauge theory of gravity.Some of these solutions have a counterpart in the crystalline theory of matter.Other solutions related to time dislocations with no analogy in solid state 
physics.Recently I presented \cite{6} a new solution representing a spherically symmetric time dislocation static solution of Einstein field equations presenting a tachyonic \cite{7} sector.
In this paper I present two new classes of solutions with spherically symmetry.At this point is important to recall that Birkoff theorem is not always valid and a vacuum solution of
Poincar\'{e} gauge gravity does not need to coincide with the Schwarzschild solution of General Relativity (GR).The first represents a time dependent solution of Einstein field equations. 
The other represents a time dependent solution obtained by the product of two homogeneous functions, one of degree two and the other of degree minus two. As it is well known this procedure yields a new function of degree minus two which again makes Cartan torsion \cite{8}to vanish.Physical effects of these two new solutions are discussed.
\section{Poincar\'{e} Direct Gauge Gravity}
\paragraph*{}
Recently Edelen \cite{1} has found solutions of Einstein type equations of a direct gauge theory of gravity of the Poincar\'{e} group $ P(10)$.In this section I shall give a brief review of this theory which will help the reader to grasp the section III.
The Minkowski spacetime $M_{4}$ with global coordinates $ \{x,y,z,t \} $ is the base of a $ L_{4} $ Riemann-Cartan Spacetime which is generated from the action of a Poincar\'{e} group on $ M_{4} $. The Riemann-Cartan manifold is , in general, endowed with both curvature and torsion. The translation group T(4) yields the compensating one-forms $ {\phi}^{i}={{\phi}^{i}}_{j}(x^{k})dx^{j} \hspace{0.5cm}(1 \le i \le 4) $ and local axial of the six-parameter L(6) local Lorentz group and ten-parameters Poincar\'{e} group P(10)$ \subset $GL(5,R)are given by
\begin{equation}
\begin{array}{ll}
{W}^{\alpha}={W^{\alpha}}_{i}(x^{k}) dx^{i} \hspace{1.5cm} (1 \le {\alpha} \le 6) \nonumber \\
{B}^{i} = {{B}^{i}}_{j}(x^{k})dx^{j}=( {{\delta}^{i}}_{j}+ {{W}^{\alpha}}_{j} {{l}^{i}}_{k \alpha} {x}^{k}+{{\phi}^{i}}_{j} )d{x}^{j} \nonumber \\
\end{array}
\end{equation}
respectively. The distortion 1-forms $ \{ B^{i} \vert 1 \le i \le 4 \}$ are the basis of a vector space $ \wedge ^{1} $ of forms on $ L_{4} $.The distorted Riemann-Cartan spacetime $ L_{4} $ obtained from $ M_{4} $ by minimal substitution yields the line element
\begin{equation}
ds^{2}=g_{ij} dx^{i} \otimes  dx^{j}
\label{2}
\end{equation}
where $ g_{ij}={{B}^r}_{i}h_{rs}{{B}^s}_{j}$ where $ g=det(g_{ij})=-B^{2} $ and $ ds^{2}=h_{ij}dx^{i} \otimes  dx^{j} $ is the $ M_{4} $ line element.
The spacetime $ L_{4} $ has both curvature and torsion in general. The Cartan torsion 2-forms $ \{ {\sum}^{i}| 1 \le i \le 4 \} $ are given by 
\begin{equation}
{\sum}^{i}=dB^{i}+W^{\alpha} {{l^{i}}_{j \alpha}} \wedge B^{j}
\label{3}
\end{equation}
Where the holonomic torsion 2-forms $                          S^{k}=\frac{1}{2}{\Gamma}_{ij}-{{\Gamma}^{k}}_{ji})dx^{i} \wedge dx^{j} $ are determined in terms of the $ \sum^{i} $ by  $ S_{k}={{b}^{k}}_{r} {\sum}^{r} $ 
where $ b_{i}  \rfloor   B^{j}={{\delta}^{j}}_{i} $ ,\linebreak $ b_{i}={{b}^{j}}_{i}(x^{k}) {\partial}_{j} $ being the frames of $ B^{j}$ .
In general the torsion forms are given by (the coframes)
\begin{equation}
{\sum}^{i}={\theta}^{\alpha} {l^{i}}_{j \alpha} {\chi}^{j} + d{\phi}^{i} + W^{\alpha} {l^{i}}_{j \alpha} \wedge {\phi}^{j}
\label{4}
\end{equation}
where $ {\theta}^{\alpha}= \frac{1}{2}{{\theta}^{\alpha}}_{rs} dx^{r} {\wedge} dx^{s} $ and the Riemann curvature is given by $ {R^{i}}_{rsj} = {{\theta}^{\alpha}}_{rs} {L^{i}}_{j \alpha} $.
In this paper we shall  be concerned with dislocations where curvature vanishes and only torsion survives.Thus $ {\theta}^{\alpha} = 0 $ , $ {R^{i}}_{rsj} = 0 $. 
Defining $ W^{\alpha} \equiv 0 $ the dislocation density and current (Cartan torsion) reduces to $ {\sum}^{i} = d{\phi}^{i} $ and the distortion 1-forms have the form $ B^{i}=dx^{i} + {\phi}^{i} $.In general in crystalline solids the procedure consists in giving the dislocation density 2-forms and then to calculate the response of the solid.Here we shall consider a dislocation density like
\begin{equation}
{\sum}^{i}= A^{i}(R,t) dR{\wedge}dt
\label{5}
\end{equation}
From the expression $ {\sum}^{i}=d{\phi}^{i} $ , $ d{\sum}^{i}= 0 $.
On integration of the system yields
\begin{equation}
{\phi}^{i}=a^{i}(R,t)(Rdt-tdR)
\label{6}
\end{equation}
the essential difference between these functions here and Edelen's functions in \cite{1} is that the functions here are not biaxial functions \cite{7}. The functions (\ref{7}) are indeed homogeneous of degree -2 outside the core of defects since Cartan torsion is
\begin{equation}
{\sum}^{i}=d{\phi}^{i}=\{ \frac{{\partial}a^{i}}{{\partial}R}R + \frac{{\partial}a^{i}}{{\partial}t}t + 2 a^{i} \} dR {\wedge} dt
\label{7}
\end{equation}
and therefore the region $ R > R_{0} $ (here $ R=\sqrt{x^{2}+y^{2}+z^{2}} $ is a homogeneous function of degree 1) if $ a^{i} $ are homogeneous of degree -2, $ {\sum}^{i}=0 $ from \ref{7} and curvature and torsion vanish. Despite of this situation the solution of Einstein field equation in vacuum $ (R_{ik}=0) $ is topologically nontrivial like the ones \cite{9} obtained earlier by Marder in the context of general relativity and by Tod \cite{8} and Letelier \cite{3,4} in the context of Einstein-Cartan theory of gravity.
\section{Space Dislocation Solution of Einstein Field Equations}
\paragraph*{}
In this section I shall be concerned with the spherically symmetric solution of Einstein field 
equation representing a space dislocation.This solution of course contains an analogy in condensed matter physics \cite{9}.The solution is obtained by choosing the $ a^{i} `s $ as 
\begin{equation}
\begin{array}{ll}
a^{1}= a^{2}=a^{4}=0 \nonumber \\
a^{3}=f(r)= \frac{K}{r^{2}} \nonumber \\
\end{array}
\end{equation}
From these expressions one obtains the following coframes
\begin{equation}
\begin{array}{lll}
B^{4}=dx^{4}=dt \nonumber\\
B^{2}=dx^{2},B^{1}=dx^{1} \nonumber\\
B^{3}=dx^{3}+ (a^{3})(rdt-tdr) \nonumber\\
\end{array}
\end{equation}
which yields
\begin{equation}
\begin{array}{ll}
B^{4}_{4}= (1+fr), B^{4}_{1}=- \frac{ftx}{r} \nonumber\\
B^{4}_{2}= - \frac{fty}{r} B^{4}_{3}=- \frac{ftz}{r} \nonumber\\
\end{array}
\end{equation}
where we have used the transformation
\begin{equation}
dr = \frac{xdx}{r}+ \frac{ydy}{r}+ \frac{zdz}{r}
\label{11}
\end{equation}
since the Poincar\'{e} group $ P(10) $ only has a representation as a subgroup of $ GL(5,R) $ when $ M_{4} $ has a rectilinear cover $ (x,y,z,t) $ modulo the homogeneous action of $ P(10) $.
The frames $ b^{i} $ read
\begin{equation}
\begin{array}{llll}
b^{4}={\partial}_{t}-\frac{fr}{\gamma}{\partial}_{z}\nonumber \\
b^{1}={\partial}_{x}-(\frac{\frac{{ft}{r}}{\gamma}}{\delta}){\partial}_{z}\nonumber \\
b^{2}={\partial}_{y}-(\frac{\frac{{ft}{r}}{\gamma}}{\delta}){\partial}_{z}\nonumber \\
b^{3}=(\frac{{\partial}_{z}}{\gamma})\nonumber \\
\end{array}
\end{equation}
where $ {\gamma} = 1 +\frac{fz}{r} $. From the metric (\ref{2}) one obtains
\begin{equation}
ds^{2}= dt^{2}-\frac{Kt^{2}}{r^{4}}dr^{2}-\frac{Kt}{r^{3}}rdrdz+\frac{K^{2}}{r^{4}}rdrdt
\label{13}
\end{equation}
Note that at $ z=constant $ hypersurfaces this metric reduces to
\begin{equation}
ds^{2} = dt^{2}-\frac{Kt^{2}}{r^{4}}dr^{2}+\frac{K^{2}}{r^{4}}rdrdt
\label{14}
\end{equation}
photons $ (ds^{2}=0) $ obey the following equations 
\begin{equation}
1=\frac{Kt^{2}}{r^{4}}(v_{r})^{2}+\frac{K^{2}}{r^{4}}rv_{r}
\label{15}
\end{equation}
If $ v_{r} $ is orthogonal to the radius r (circular orbits) the last term in {\ref{15}} vanishes and we are left with
\begin{equation}
v_{r}=\frac{r^{2}}{K^{\frac{1}{2}}t}
\label{16}
\end{equation}
which vanishes for as time t goes to infinite as long as the photon is bound to a certain finite radius.The same result can be 
obtained for a bound time and a vanishing radius.This can be interpreted as saying that the photon will fall back towards the core defect. As happened before in the Poincar\'{e} gauge gravity
developed by Edelen \cite{2}, tachyons are also present here as can be seen as follows. Since 
$ V^{i} = b^{i}_{j} v^{j} $ and $  v^{k} =(0,k,0,0) $,k being a constant; the velocity of 
test particle is given by
\begin{equation}
{V}^{2} = {k}^{2}( 1 + \frac{{K}^{2}}{{k}^{2}}({X}^{2} + {k}^{2}{\tau}^{2}))
\label{17}
\end{equation}
where $ {\tau} $ is the proper time. The geodesic motion is given by $ t={\tau} $ and   
\begin{equation}
\begin{array}{ll}
\frac{dx}{d{\tau}}=0 , \frac{dy}{d{\tau}}=k \nonumber \\
\frac{dz}{d{\tau}}=- \frac{k}{X} \nonumber
\end{array}
\end{equation}
At $ {\tau}=0 $ the velocity (\ref{17}) along the z-direction yields
\begin{equation}
V_{z}= \frac{dz}{d{\tau}} = - \frac{K}{x^{2}+z^{2}} < 0
\label{18}
\end{equation}
Implying that the particle is falling down and is constant at $ z={0} $ plane.From (\ref{18})
around $ {\tau}=0 $ is clear that 
\begin{equation}
V^{2}= - {k}^{2}( 1 + \frac{{K}^{2}}{{X}^{2}}) > {k}^{2}
\label{19}
\end{equation}
Since $ V^{4}= \frac{dt}{d{\tau}}=1 $ the 4-velocity measured by a Minkowski observer reads
\begin{equation}
{V}^{2} = {V}^{i}{g}_{ij} {V}^{j} = ( {V}^{4})^{2} - {V}^{2} = 1 - {k}^{2}( 1 + \frac{\frac{K^{2}}{k^{2}}}{{X}^{2} + {k}^{2}{\tau}^{2}})
\label{20}
\end{equation}
At $ {\tau}=0  $ this reduces to 
\begin{equation}
{V}^{2} = 1 - {k}^{2}( 1 + \frac{{K}^{2}}{{k}^{2}{X}^{2}})
\label{21}
\end{equation}
and the spatial part is given by (\ref{21}) shows that there is a tachyonic sector,since $V^{2}>k^{2}$. Thus our solution is a spherically symmetric vacuum solution of the Direct Poincar\'{e} 
gauge gravity. In the next section I shall present a time dependent solution generated by the time independent solution given in this section.
\section{Time Dependent Spherically Symmetric Solution}
\paragraph*{}
It is well known that the product of a homogeneous function of degree zero by an homogeneous function of degree minus two generates an homogeneous function of degree minus two.In this 
section I shall make use of this reasoning to generate a new spherically symmetric solution of Einstein type field equations of a gauge theory of gravity.Let us consider the ansatz
\begin{equation}
f(r,t)=(\frac{K}{r^{2}})ln(\frac{r}{t})
\label{22}
\end{equation}
This solution is easily seen to fulfill the homogeneity condition discussed above. The solution produced by the expression (\ref{22}) is a time dependent torsion-free spherically symmetric 
solution of Poincar\'{e} gauge gravity.Let us now follow the same steps of the previous section to
investigate the presence of tachyons in the theory.From eqns. (\ref{14}) and 
$ b_{i} = b^{j}_{i}{\partial}_{j} $ one obtains the following expressions
\begin{equation}
\begin{array}{lll}
b^{4}_{4}=1, b^{3}_{4}= -(\frac{K}{r})(\frac{ln{\alpha}}{\beta})\nonumber\\
(b^{1})_{1}=1 ,(b^{3})_{1}=-{\delta} \nonumber\\
(b^{2})_{2}=1 ,(b^{3})_{2}=-{\delta} \nonumber\\
\label{23}
\end{array}
\end{equation}
where $ {\delta}=- \frac{Kt ln{\alpha}}{r^{2}{\beta}}$ and $ {\beta} = (1 + \frac{Kz ln{\alpha}}{r^{2}}) $ and $ {\alpha}=\frac{r}{t} $. Choosing once more $ v^{k}=(0,k,0,1) $ and by making use of (\ref{23}) one may compute the velocity $ V^{k} $ as
\begin{equation}
\begin{array}{llll}
V^{4}=1 \nonumber\\
V^{1}=0 \nonumber\\
V^{2}=k \nonumber\\
V^{3}= - ([ 1 + \frac{k{\tau}}{r}]) \frac{Kln{\alpha}}{\beta}  \nonumber\\
\end{array}
\end{equation}
The expression $ V^{3} $ can be approximatly written as
\begin{equation}
V^{3} = -K \frac{ln r}{r} - K \frac{lnz}{r}
\label{25}
\end{equation}
as r tends to infinite. At very large distances from the core defect $ r >> r_{0}$, where $ r_{0} $ is the core radius, and for bound proper times one observes that $ V^{3} $ reduces to
\begin{equation}
V^{3}=-K \frac{lnr}{r}
\label{26}
\end{equation}
and the velocity of the photon along the z-direction is bound by the gravitational attraction of the core defect.Notice that formula (\ref{26}) also  tell us that the initially nonstatic 
configuration tends to a static dislocation as time goes by.This is a typical behaviour for a nonstatic gravitational collapse.The model discussed here it is very similar to the model of 
the geodesic motion of the electrons around dislocation as discussed previously by F.Moraes \cite{10} since our photons move also along geodesics around dislocations.Investigation of 
autoparalles around gravitational dislocation are now under investigation.Geodesics and autoparallels around other spacetime defects \cite{11,12} can also be investigated in near future.
\section*{Acknowledgments}
\paragraph*{}
I am very grateful to Prof.Dominic Edelen for providing fundamental ideas for the developementof this work.Thanks are also due to Fernando Moraes,Axel Pelster for helpful discussions on the subject of this paper.Thanks are due to CNPq. and DAAD(Bonn) from financial support.Finally I would like to thank 
Prof.H.Kleinert for his kind hospitality at the Institut fur Theoretische Physik,Freie Universitat,Berlin.

\end{document}